\documentstyle[12pt]{article}
\normalbaselines
\title{Discontinuity surface instead of singularity}
\author{\bf M. V. Gorbatenko\footnote{Russian Federal Nuclear Center - VNIIEF; Sarov, Nizhni Novgorod Region, 607190, Russia; e-mail: gorbatenko@vniief.ru}}
\date{}
\begin{document}
\maketitle

\begin{abstract}
Einstein equations are addressed with the energy-momentum tensor that 
appears if the equations under discussion are required to possess conformal 
invariance. It is proved that thus derived equations (equations of 
conformally invariant geometrodynamics) can have not only smooth solutions, 
but also solutions with discontinuities on space-like hypersurfaces. The 
solutions obtained are similar to the well-known discontinuous Einstein 
equation solutions like shock-wave solutions, extended-source solutions, 
etc.

For the centrally symmetric stationary solution discussed in the paper, the 
discontinuity surface removes the singularity. The degree of generality of 
this solution regularization mechanism is discussed.

The issue of the mechanism that forces any smooth solution in the 
conformally invariant geometrodynamics to be rearranged into the 
discontinuous one when certain conditions are met is also discussed. The 
conditions can be: (1) sound speed becoming to be higher than light speed; 
(2) the solution becoming intolerant to smaller and smaller-scale 
perturbation modes.
\end{abstract}

\vskip20mm

This paper is a summary of ref. [1] that contains missing proofs of the 
statements presented here and needed references. 

We consider discontinuous solutions to Einstein equations with the nonzero 
energy-momentum tensor. The equations are

\begin{equation}
\label{eq1}
R_{\alpha \beta}  - {\textstyle{{1} \over {2}}}g_{\alpha \beta}  R = 
T_{\alpha \beta}  ,
\end{equation}

\noindent
where

\begin{equation}
\label{eq2}
T_{\alpha \beta}  = - 2A_{\alpha}  A_{\beta}  - g_{\alpha \beta}  A^{2} - 
2g_{\alpha \beta}  A^{\nu} _{;\nu}  + A_{\alpha ;\beta}  + A_{\beta ;\alpha 
} + g_{\alpha \beta}  \lambda ,
\end{equation}

 $A_{\alpha}  $ is a so-called gauge vector, $\lambda $ is, generally speaking, 
one of the desired functions. Equations (\ref{eq1}) retain their form in conformal 
transformations

\begin{equation}
\label{eq3}
g_{\alpha \beta}  \to g_{\alpha \beta}  \cdot e^{2\sigma} ,\quad A_{\alpha}  
\to A_{\alpha}  - \sigma _{;\alpha}  ,\quad \lambda \to \lambda \cdot e^{ - 
2\sigma} .
\end{equation}

Therefore we refer to equations (\ref{eq1}) with energy-momentum tensor (\ref{eq2}) as 
equations of conformally invariant geometrodynamics.

The Riemannian space complemented with $A_{\alpha}  $, $\lambda $ and 
transformation rules (\ref{eq3})can be considered as Weyl space. In the Weyl space 
the connectivity $\Gamma _{\alpha}  ^{\lambda} {} _{\beta}  $ is expressed in 
terms of Christoffel symbols $\left( {_{\alpha \beta} ^{\lambda} }  \right)$ 
and vector $A_{\alpha}  $ with relation

\begin{equation}
\label{eq4}
\Gamma _{\alpha}  ^{\lambda} {}_{\beta}  = \left( {_{\alpha \beta} ^{\lambda 
}}  \right) + \delta _{\alpha} ^{\lambda}  A_{\beta}  + \delta _{\beta 
}^{\lambda}  A_{\alpha}  - g_{\alpha \beta}  g^{\lambda \tau} A_{\tau}  .
\end{equation}

Equations (\ref{eq1}) with energy-momentum tensor (\ref{eq2}) possess a unique property: 
Cauchy's problem for these equations is posed without any connections to 
Cauchy's data. Thus, in the synchronous frame $\left( {g_{0k} = 0} \right)$ 
with using gauge condition $\lambda = Const$ the Cauchy's data are the 
following sixteen functions: $g_{mn} ,\;\dot {g}_{mn} ,\;g_{00} ,\;A_{k} $. 

The starting axiomatics of the Weyl space requires that the minimum 
necessary geometric object smoothness classes specified in Table 1 be 
ensured.

Table 1 - Minimum smoothness classes

\bigskip

\noindent
\begin{center}
\begin{tabular}{|p{155pt}|p{106pt}|p{113pt}|}\hline
Object& 
Smoothness class as a whole on the manifold& 
Smoothness class on local charts (pieces) \\
\hline
Components of metrics $g_{\alpha \beta}  $& 
$C^{1}$
&$C^{2}$ \\
\hline
Christoffels $\left( {_{\alpha \beta} ^{\lambda} }  \right)$& 
$C^{0}$&$C^{1}$ \\
\hline
Components of gauge vector$A_{\alpha}  $& 
Piecewise continuous& 
$C^{0}$ \\
\hline
Function $\lambda $& 
Piecewise continuous& 
$C^{0}$ \\
\hline
\end{tabular}
\end{center}

In this paper our interest is not with light-like discontinuity surfaces, 
but with space-like hypersurfaces. Such discontinuity surfaces are known to 
appear only when the energy-momentum tensor is nonzero. These solutions 
include gas-dynamic shock-wave type solutions, extended-source solutions, 
etc. What follows will be analogous to the shock wave theory within the 
framework of the general relativity theory, but under the sole condition: 
the energy-momentum tensor will be of form (\ref{eq2}). 

Apply the requirement of the existence of only minimum smoothness classes to 
the static centrally symmetric problem. The squared integral for this 
solution, without loss of generality, can be reduced to

\begin{equation}
\label{eq5}
ds^{2} = - exp\left( {\gamma}  \right) \cdot dt^{2} + exp\left( { - \gamma}  
\right) \cdot dz^{2} + exp\left( {\beta}  \right) \cdot \left( {d\theta ^{2} 
+ sin^{2}\theta \cdot d\varphi ^{2}} \right)
\end{equation}

\noindent
and the gauge vector to

\begin{equation}
\label{eq6}
A_{\alpha}  = \left( {\phi ,0,0,0} \right).
\end{equation}

For four desired functions $\gamma ,\;\beta ,\;\phi ,\;\lambda $ relations 
(\ref{eq1}) result in four equations which we will not present here. Instead, we at 
once present the general solution branch, which can transfer to de Sitter 
solution and Schwartzschield solution.

\begin{equation}
\label{eq7}
\left. {\begin{array}{l}
 {\phi = p \cdot exp\left( {\gamma}  \right)} \\ 
 {exp\left( {\beta}  \right) = A \cdot sh^{2}\left( {pz + a} \right)} \\ 
 {exp\left( {\gamma}  \right) = \frac{{1}}{{p^{2}A}} + B \cdot \left[ {pz 
\cdot cth\left( {pz + a} \right) - 1} \right] + bp \cdot cth\left( {pz + a} 
\right)} \\ 
 {\lambda \left( {z} \right) = Bp^{2}} \\ 
 \end{array}}  \right\}
\end{equation}

Seek the solution, such that has a discontinuity for radial variable $z = 
\Delta $. In region $z < \Delta $ the desired solution is a de Sitter type 
solution, while in region $z > \Delta $ it is an approximation to the 
Schwartzschield solution. Denote the constants that describe the desired 
solution in both the regions as

\begin{equation}
\label{eq8}
\left\{ {\begin{array}{l}
 {p_{0} ;\;\;A_{0} ;\;\;a_{0} = 0;\;\;B_{0} ;\;\;b_{0} = 0\quad \quad z < 
\Delta}  \\ 
 {p;\;\;A = \frac{{1}}{{p^{2}}};\;\;a;\;\;B = 0;\;\;b\quad \quad \quad \quad 
z > \Delta}  \\ 
 \end{array}}  \right.
\end{equation}

The lacing conditions follow from Table 1 and reduce to

\begin{equation}
\label{eq9}
\left. {\left. {\left[ {e^{\beta} } \right]} \right|_{\Delta}  = \left. 
{\left[ {e^{\beta} {\beta} '} \right]} \right|_{\Delta}  = \left[ {e^{\gamma 
}} \right]} \right|_{\Delta}  = \left. {\left[ {e^{\gamma} {\gamma} '} 
\right]} \right|_{\Delta}  = 0,
\end{equation}

\noindent
that is to ensuring continuity on the discontinuity surfaces of functions \\
$\gamma ,\quad {\gamma} ',\quad \beta ,\quad {\beta} '$. Under assumptions 
(\ref{eq8}) made, conditions (\ref{eq9}) determine four of seven problem parameters $p_{0} 
;\;A_{0} ;\;B_{0} ;\;\Delta ;\;p;\;b;\;a$. 

In what follows it would be convenient to introduce parameter $z_{0} \equiv 
- a/p$ in lieu of $a$ and use not parameters $p_{0} ;\;A_{0} ;\;B_{0} 
;\;\Delta ;\;p;\;b;\;z_{0} $ themselves, but their dimensionless 
combinations with retaining the old notations for the dimensionless 
parameters (excluding $z_{0} $),

\begin{equation}
\label{eq10}
p_{0} \to p_{0} \Delta ;\quad p \to p\Delta ;\quad A_{0} \to \frac{{A_{0} 
}}{{\Delta ^{2}}};\quad b \to \frac{{b}}{{\Delta} };\quad \varepsilon \to 
\frac{{z_{0}} }{{\Delta} }.
\end{equation}

Upon the transition to the dimensionless parameters the number of the 
parameters becomes six. Thus, the problem of lacing possesses the 
self-similarity.

Assume that constants $p_{0} $ and $A_{0} p_{0}^{2} $ are

\begin{equation}
\label{eq11}
p_{0} = 2.77,\quad A_{0} p_{0}^{2} = 6.94376 \quad .
\end{equation}

From conditions (\ref{eq9}) we find:

\begin{equation}
\label{eq12}
p = 2.7887,\quad \varepsilon = 0.954694,\quad B_{0} = 2.8803 \cdot 10^{ - 
4},\quad b = - 0.305319.
\end{equation}

Hence, the accurate solution to the centrally symmetric static problem with 
the discontinuity surface has been constructed, such that the solution is an 
analog of the de Sitter solution on the left of the surface and of the 
Schwartzschield solution on the right. 

Energy-momentum tensor (\ref{eq2}) can be analyzed by the well-known Eckart method. 
In gauge

\begin{equation}
\label{eq13}
\lambda = Const,\quad A^{\mu} _{;\mu}  = 0,
\end{equation}

\noindent
which is valid in the centrally symmetric static problem as well, the 
current density vector is written as

\begin{equation}
\label{eq14}
j_{\alpha}  = - 2\lambda A_{\alpha}  .
\end{equation}

In the geometrodynamics scheme under discussion, the retained substance 
density $\rho $ is related to vector $j_{\alpha}  $ as

\begin{equation}
\label{eq15}
j^{\alpha}  = \rho \cdot u^{\alpha} .
\end{equation}

In the centrally symmetric problem,

\[
u^{\alpha}  = \left( {exp\left( { - \gamma /2} \right);0;0;0} \right);\quad 
\quad u_{\alpha}  = \left( { - exp\left( {\gamma /2} \right);0;0;0} 
\right).
\]

The determined energy density $U$ and pressure $P$ result in the following 
expressions for these quantities:

\begin{equation}
\label{eq16}
 - U \equiv T_{0}^{0} = 3exp\left( { - \gamma}  \right) \cdot \phi ^{2} + 
\lambda .
\end{equation}

\begin{equation}
\label{eq17}
P \equiv \frac{{1}}{{3}}\left( {T_{1}^{1} + T_{2}^{2} + T_{3}^{3}}  \right) 
= exp\left( { - \gamma}  \right) \cdot \phi ^{2} + \lambda .
\end{equation}

From (\ref{eq16}), (\ref{eq17}) it follows that for the static centrally symmetric problem 
the equation of state of geometrodynamic medium is

\begin{equation}
\label{eq18}
P = \left( { - 5/9} \right)U.
\end{equation}

Assume that for the geometrodynamic medium the law of degradation of energy 
is valid in the form, in which it takes place for viscous heat-conducting 
continuum. This assumption leads to Maxwell cross relation

\begin{equation}
\label{eq19}
\frac{{\partial} }{{\partial V}}\left( {\frac{{\partial S}}{{\partial P}}} 
\right) = \frac{{\partial} }{{\partial P}}\left( {\frac{{\partial 
S}}{{\partial V}}} \right).
\end{equation}

Having solved the differential relation appearing from (\ref{eq19}), we arrive at 
the relation of temperature $T$ to specific volume:

\begin{equation}
\label{eq20}
T = \frac{{Q_{0}} }{{V^{1/3}}} = Q_{0} \left( {\frac{{8}}{{3}}} 
\right)^{1/6}\sqrt {\lambda}  .
\end{equation}

(here $Q_{0} $ is some dimensional constant), and entropy density $S$ is

\begin{equation}
\label{eq21}
S = \frac{{3^{2/3}}}{{2 \cdot Q_{0} \lambda} } + S_{0} .
\end{equation}

In the stationary centrally symmetric problem under discussion in the region 
gauge (\ref{eq13}) is used in all the thermodynamic quantities are constant and 
expressed in terms of $\lambda $. 

The isentropic sound speed for the internal part of the solution is

\begin{equation}
\label{eq22}
c_{s}^{2} = \frac{{7}}{{3\sqrt {6\lambda} } }.
\end{equation}

From the condition of equal sound speed and light speed we obtain relation

\begin{equation}
\label{eq23}
c_{s}^{2} = 8\pi r_{0} .
\end{equation}

From (\ref{eq22}), (\ref{eq23}) it follows that the criterion value $\lambda _{cr} $ of 
$\lambda $, on whose achievement the discontinuity surface appears, is

\begin{equation}
\label{eq24}
\lambda _{cr} = \frac{{7^{2}}}{{2^{7} \cdot 3^{3} \cdot \pi ^{2} \cdot 
r_{0}^{2}} }.
\end{equation}

The value of $\lambda = \lambda _{cr} $ is established as a result of some 
dynamic transient, whose description is beyond the scope of this paper. The 
transient apparently proceeds until $\lambda $ in the internal part of the 
solution rises up to a level, at which the sound speed becomes equal to the 
light speed.

The $\lambda $ is related to the parameters (\ref{eq10}) as $\lambda = B_{0} 
p_{0}^{2} $ and the $r_{0} $ is nothing else but $r_{0} = - b$. Relation 
(\ref{eq24}) for the solution under discussion is therefore equivalent to

\begin{equation}
\label{eq25}
B_{0} p_{0}^{2} b^{2} = \frac{{7^{2}}}{{2^{7} \cdot 3^{3} \cdot \pi ^{2}}}.
\end{equation}

Thus, the condition of equal sound speed and light speed leads to the 
relation between the two independent constants $p_{0} $, $A_{0} p_{0}^{2} $. 
The dependence of $A_{0} p_{0}^{2} $ on $p_{0} $ following from condition 
(\ref{eq25}) is broken at $p_{0} = 2.77$, at higher $p_{0} $ equation (\ref{eq25}) has no 
solutions. In the range of the values of $p_{0} $, at which equation (\ref{eq25}) is 
solvable, the curve $ - b\left( {p_{0}}  \right)$ monotonically decreases 
and reaches its minimum at the boundary value $p_{0} = 2.77$. In so doing 
$A_{0} p_{0}^{2} = 6.94376$, $b = - 0.305319$. It is these boundary values 
of the constants $p_{0} $, $A_{0} p_{0}^{2} $ that are used in (\ref{eq11}). 

The existence of the minimum in $ - b\left( {p_{0}}  \right)$ means that the 
solution with constants (\ref{eq11}) possesses stability in the following sense. If 
there were no minimum, the parameter $ - b\left( {p_{0}}  \right)$ could 
decrease to zero; as the parameter is equal to the Schwartzschield radius, 
the vanishing would essentially mean the mass ``vanishing''. 

With gauge (\ref{eq13}), the dynamic coefficient of viscosity is

\begin{equation}
\label{eq26}
\eta = \frac{{\rho} }{{2\lambda} }.
\end{equation}

In viscous medium, the solution evolves according to the dynamic equations 
until Reynolds number $R$ reaches a critical value $R_{0} $ leading to 
appearance of instability of some motion modes, that is turbulence. 

Pose the problem to find size $L_{0} $, beginning with which the Reynolds 
number reaches the $R_{0} $. From the definition of the Reynolds number it 
follows that with the system size $L_{0} $, characteristic medium velocity 
$u$, and mass $m$ observed from the outside the following relation takes 
place:

\begin{equation}
\label{eq27}
L_{0} = \frac{{R_{0} c^{3}}}{{u \cdot 16\pi G\lambda m}}.
\end{equation}

To find $L_{0} $, as it follows from (\ref{eq27}), some assumptions of the value of 
$u$ as well as an assumption allowing $\lambda $ to be estimated have to be 
made. 

As for $u$, assume that it characterizes the perturbation velocity, that is 
take

\begin{equation}
\label{eq28}
u = c.
\end{equation}

In view of this assumption, relation (\ref{eq27}) becomes

\begin{equation}
\label{eq29}
L_{0} = \left( {\frac{{R_{0}} }{{16\pi} }} \right) \cdot \frac{{1}}{{\lambda 
r_{0}} } \quad ,
\end{equation}

\noindent
where $r_{0} $ is the gravitational radius of the object. Assume that the 
coefficient in brackets is on the order of one. Then from (\ref{eq29}) it follows 
that

\begin{equation}
\label{eq30}
L_{0} \sim\frac{{1}}{{\lambda r_{0}} }.
\end{equation}

Find the characteristic size $L_{0} $ for the above-constructed 
discontinuous stationary centrally symmetric geometrodynamic solution. We 
obtain

\begin{equation}
\label{eq31}
L_{0} \sim\frac{{2^{7} \cdot 3^{3} \cdot \pi ^{2} \cdot r_{0}} }{{7^{2}}} 
\approx 120 \cdot r_{0} .
\end{equation}

Thus, the intolerance of the centrally symmetric solution to perturbations 
can appear at the values of the radial variable $L_{0} $ much higher than 
the gravitational radius. However, its specific value depends on the choice 
of the Reynolds number for the criterion value.

Hence, equations (\ref{eq1}) with tensor $T_{\alpha \beta}  $ of form (\ref{eq2}) admit the 
existence of discontinuous solutions. The possibility in itself to construct 
the discontinuity-surface solution is a nontrivial fact. Note that, for 
example, the geometrodynamic equations without $\lambda $ term do not admit 
construction of the solution obtained in the paper.

The discontinuous solution constructed is regular in the entire radial 
variable range, which the geodesic completeness takes place in. What makes 
the solution to be rearranged with the transition of the Schwartzschield 
branch to the de Sitter branch at certain radial variable? The question 
touches upon the fundamental problem of the unlimited energy cumulation 
instability, that is the problem addressed in many papers. We do not have a 
full answer to the question, at least, for the reason that in this paper we 
restrict our consideration to the scope of the static centrally symmetric 
problem. However, we can specify the peculiar geometrodynamics features that 
can lead to the singularity removal by the discontinuity surface formation. 

First, in the scheme under discussion the space-time evolution proceeds in 
accordance with the dynamic equations at any time. No manual introduction of 
equations of state or coefficients of viscosity is admitted. As a result, 
meeting the energy dominance condition, which collapses are typically 
associated with, is not guaranteed whatsoever.

Second, with increasing $\lambda $ viscosity decreases, Reynolds number 
increases. When the Reynolds number has reached some criterion value $R_{0} 
$, the solution becomes intolerant to the perturbations of characteristic 
size (\ref{eq31}). It is hard to tell as applied to geometrodynamic medium at what 
specific value of $R_{0} $ the perturbations will rearrange the solution. 
But that the solution rearrangement is inevitable follows from the fact that 
smaller and smaller-scale perturbations become unstable as the singularity 
is approached. At some stage the turbulization will come to be caused by 
quantum fluctuations of vacuum.

The above-mentioned features of the dynamic equations allow us to suggest 
that the mechanism of the solution regularization in geometrodynamics 
through the discontinuity surface formation can be of general nature. This 
paper has validated this suggestion by the specific example. 

The work was carried out under support in part by the International Science 
and Technology Center (Project \# KR-677).

\section*{References}

\bigskip

[1] M.V. Gorbatenko. \textit{Voprosy Atomnoi Nauki i Tekhniki}. Seriya: 
Teor. i Prikl. Fizika. \textbf{1-2}, 9-21 (2002) [In Russian].

\end{document}